\documentclass[article,aip,pof,
reprint,
 amsmath,amssymb
]{revtex4-1}

\usepackage{graphicx}
\usepackage{dcolumn}
\usepackage{bm}
\usepackage{color}
\newcommand{\W}{8.5cm}
\newcommand{\cor}[1]{\textcolor{black}{#1}}

\begin{document}

\preprint{APS/123-QED}

\title{Stokes flow paths separation and recirculation cells in X-junctions of varying angle between 
straight channels.}

\author{M. Cachile}\email{mcachil@fi.uba.ar}
\affiliation{Grupo de Medios
Porosos, Facultad de Ingenier\'{\i}a, Paseo Colon 850, 1063, Buenos
Aires (Argentina)}
\author{L. Talon}
\affiliation{Univ Pierre et Marie Curie-Paris 6, Univ Paris-Sud, CNRS, F-91405.
  Lab FAST, B\^at 502, Campus Univ, Orsay, F-91405 (France).}
\author{J.M. Gomba}\email{jgomba@exa.unicen.edu.ar}
\affiliation{IFAS, UNCPBA, Tandil (Argentina).}
\author{J.P. Hulin}\email{hulin@fast.u-psud.fr}
\affiliation{Univ Pierre et Marie Curie-Paris 6, Univ Paris-Sud, CNRS, F-91405.
  Lab FAST, B\^at 502, Campus Univ, Orsay, F-91405 (France).}
\author{H. Auradou}\email{auradou@fast.u-psud.fr}
\affiliation{Univ Pierre et Marie Curie-Paris 6, Univ Paris-Sud, CNRS, F-91405.
  Lab FAST, B\^at 502, Campus Univ, Orsay, F-91405 (France).}


\date{\today}

\begin{abstract}
Fluid and solute transfer in X-junctions between straight channels is shown   to depend critically  on the junction angle $\alpha$ in the Stokes flow regime.
Experimentally, water and a water-dye solution
are injected at equal flow rates  in two facing  channels of the  junction: \cor{Planar Laser Induced Fluorescence (PLIF) measurements} show
that the largest part of each injected fluid ``bounces back'' preferentially into the outlet channel 
at the lowest angle to the injection; this is opposite to the inertial case and requires a high curvature of the corresponding streamlines.  \cor{The proportion of this fluid in the other channel decreases from 50\% at $\alpha = 90^{\circ}$ to zero at a threshold angle.} 
These counterintuitive features reflect the minimization of 
 energy dissipation for Stokes flows. 
Finite elements numerical simulations of a $2D$ Stokes flow of equivalent geometry confirm these results and
 show that,  \cor{below the threshold angle $\alpha_c = 33.8^\circ$}, 
 recirculation cells are present in the center part of the 
 junction and  separate the two injected flows of the two solutions. Reducing further $\alpha$ leads to the appearance
of new recirculation cells with lower  flow velocities.
\end{abstract}

\maketitle
Flow control in microchannels  has 
become an important  area of research in microprocess engineering \cite{Nguyen2005}. 
The behavior of the fluids at junctions in microfluidic circuits is particularly
critical for  applications such as mixing,  chemical reactions or heat exchange: this is for instance the
 case when two different fluids are injected separately into a junction of flow channels. While, for
T-junctions,  the flow of  the two fluids at the outlet depends 
weakly on the junction angle at low Reynolds numbers~\cite{Thomas2010a,Thomas2010b}, there  are very diverse 
 patterns  in X-junctions. Then, the  distribution of the  injected fluids between
 the two outlets and their interaction within the junction,  which are both crucial for applications,
    depend strongly on the junction angle $\alpha$ ($0 < \alpha \le 90^\circ$).
    
   For this reason, while most previous studies  dealt only with orthogonal~\cite{Hellou2011} or  
 parallel~\cite{Cochrane1981,Jeong2001} channels, the present work analyzes  specifically  the influence of the angle  $\alpha$.  More precisely, the two fluids are injected in facing channels and we study as a function of $\alpha$ their  relative fraction in each of the outlet channels and the appearance of viscous eddies. Other authors  used  tangent channels at a varying angle~\cite{Lee2009} but  the interaction between the streams differs strongly from the present case of channels in a same plane.
This system is studied by  combining experimental $PLIF$ measurements on 
 transparent models and $2D$ numerical simulations.  
   Like for  Fan and Hassan~\cite{Fan2010}, the experimental system is upscaled in order  to visualize conveniently the flow; 
however,  the Reynolds number $Re = \rho aU/\eta$ is  low enough so that inertial effects are negligible and
the Stokes equation is satisfied ($U$, $\rho$ and $\eta$ are respectively the  mean
velocity, density, and viscosity of the fluid and $a$ is the width of the channels). The flow velocity field is therefore the same 
as in much smaller microchannels of same geometry.
 
\begin{figure}[htbp]
\includegraphics[width=\W]{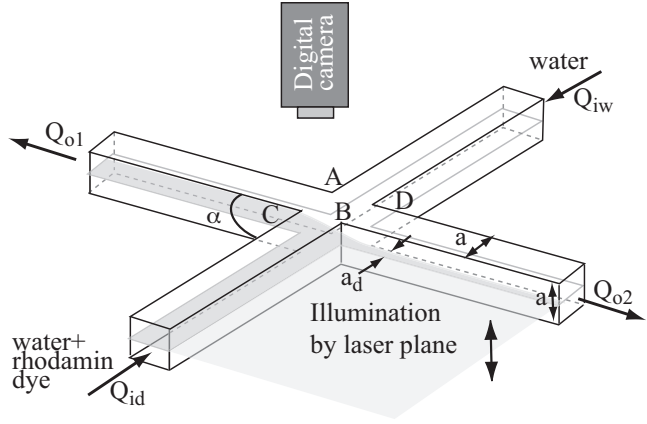}
\caption{\label{fig:fig0}Experimental setup.}
\end{figure}
The channels are carved into a transparent plexiglas plate by a computer controlled
milling machine; both their width $a$ and depth $h$ are constant and
equal to $3.55\ mm$. The horizontality of the plate is carefully adjusted. Each outlet
 is connected to one of the inlets of a double syringe pump sucking the fluids at equal  
 flow rates  $Q_{o1}$ and $Q_{o2}$ (Fig.\ref{fig:fig0}). The inlet channels $iw$ and $id$ 
 are connected to two glass flasks containing respectively pure water
 and a solution of water  and rhodamin 6G dye ($0.25\ \mathrm{mg.l}^{-1}$). 
 Due to this low concentration, the two fluids are Newtonian:
 their viscosities and densities are so close when their temperatures are matched
 that no buoyancy or viscosity driven instabilities may appear. 
Each bottle is placed on computer controlled scales allowing for
a measurement of the flow rates $Q_{iw}$ and $Q_{id}$ in the corresponding   outlet.
In the  present work, these flow  rates are set to be equal within $\pm 0.3\,\%$ by 
adjusting the relative levels of the fluid in the flasks. 

\begin{figure}[htbp]
\includegraphics[width=\W]{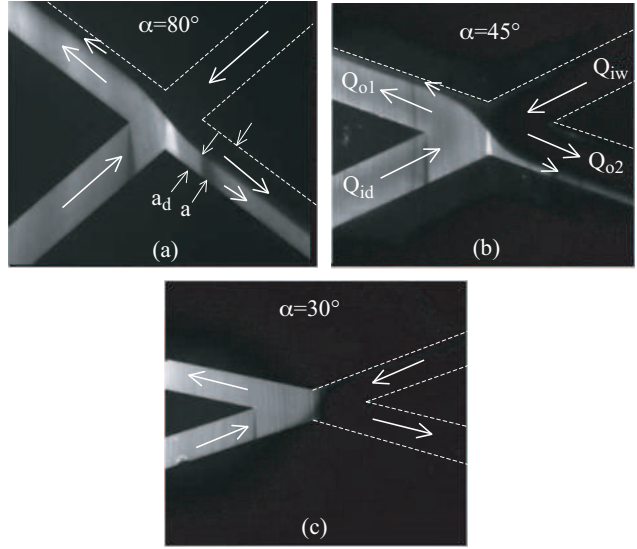}
\caption{\label{fig:fig1} Experimental images  \cor{displaying}  the distribution of the fluids in the junction
for $Re = 5$. (a) :  $\alpha=80^\circ$), (b): $\alpha=45^\circ$, (c): $\alpha=30^\circ$.}
\end{figure}

The distribution of the two fluids in the intersection is visualized
by the Planar Laser Induced Fluorescence (PLIF) technique.
A plane horizontal green laser sheet of wavelength $532\, \rm{nm}$ and thickness $\simeq 1\, \rm{mm}$, 
 parallel to both channels, enters the plexiglas model from the side.
The parts of the channels containing the rhodamin dye solution are easily identified by the emission of a
 yellow fluorescence  light of intensity  proportional to the local dye concentration.
The distribution of this solution in the illuminated plane is  then recorded by a digital camera 
located above the model :  unwanted reflected light from the laser is eliminated by a narrow bandwidth 
notch filter corresponding to its wave length $\lambda = 532 \pm 25\, \rm{nm}$ and placed in front of the 
camera sensor. Regions in which only water is present
appear as dark while those containing rhodamin are illuminated (Fig.~\ref{fig:fig1}).
\cor{In the present work, our analysis is based on the geometry of the boundary between the zones occupied 
by the two fluids.}
The laser is mounted on a vertical  translation stage: this allowed us  
 to scan the distribution of the fluids at different heights 
 $z$ for several Reynolds numbers  ($0.5 \le Re \le 50$). 
 
\cor{The distributions 
of the fluids in the symmetry plane at different $Re$ values  have first been compared
in order to check the validity of the linear Stokes equation.
For $Re = 0.5$ and $Re = 5$, the geometry of the boundary between the transparent 
 and dyed fluid is  the same: this implies that the linearity condition is satisfied and 
 that the flow is purely viscous. 
 At higher $Re$ values  ($Re \gtrsim 15$), inertial effects induce, as expected, changes of  
 the shape of the boundary.}  
Moreover, for $Re = 0.5$ and $Re = 5$,  there is no visible influence of the
vertical coordinate $z$ of the light sheet:  the distribution of the two fluids can
 then   be considered  as bidimensional. 
The flow field itself is however not bidimensional because of the vertical velocity gradients 
created by the  upper and lower wall:  however, the orientation of the velocity
and, therefore, the streamlines are approximately invariant with $z$. 
In contrast, for $Re \gtrsim 40$,   $3D$ structures appear in the fluid distribution.
 \cor{The boundary between the dyed and the transparent fluid in the images is  fuzzier
for $Re = 0.5$ than for $Re = 5$ and the thickness of the transition zone increases with 
distance to a value  of the order of $0.5\, \mathrm{mm}$ for a path length of the order of a cm .
 This value is of the same order 
of magnitude as that corresponding to transverse molecular diffusion 
 (using $D_m  = 2.8\, 10^{-10}\,m^2.s^{-1}$ at $T = 22^{\circ} C$~\cite{Gendron2008}).} 

Taking into account the above results, all experiments reported below were performed at $Re = 5$
for which  both sharp boundaries and Stokes flow are obtained.
Cells with $6$ different junctions angles $\alpha = 90, 80, 45, 30, 20$ and $10^\circ$ have been used.
Figure~\ref{fig:fig1} displays pictures obtained  in the stationary regime for three
angles $\alpha = 80^\circ$, $45^\circ$ and $30^\circ$.
The dark and illuminated regions of the junction are symmetrical with respect 
 to the center of symmetry of the junction :   this  confirms that the  flow rates 
 $Q_{id}$ and $Q_{iw}$  at the two inlets are  equal.
A first key observation is that, for all angles $\alpha$,
each injected fluid flows dominantly towards the outlet at the 
lowest angle with the injection channel (outlet $1$ for the fluorescent solution):
 this  corresponds 
to flow paths with a higher curvature  than those leading to the  other outlet.
 The  flow of each fluid is  equally split between the outlets only for  
 $\alpha = 90^\circ$. 
At $\alpha = 45^\circ$, the fraction of \cor{dyed} fluid moving towards outlet $2$
is  small and  is  exactly zero  for $\alpha = 30^\circ$.
In this latter case, all incoming streamlines ``bounce back''  
on the junction  towards outlet $1$. The main flows of the two fluids are then 
completely separated. 

\begin{figure}[htbp]
\includegraphics[width=\W]{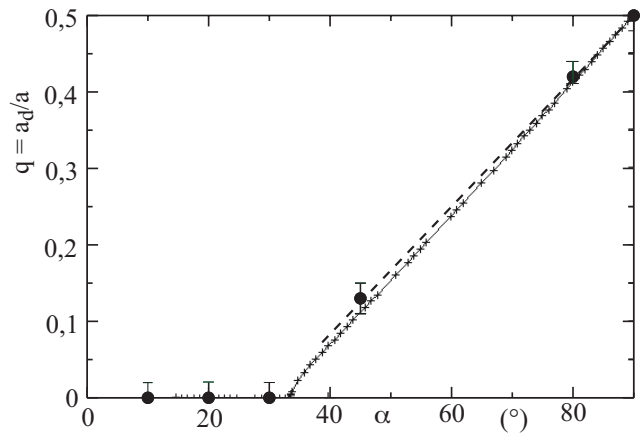}
\caption{\label{fig:fig2}Variation of ratio $q = a_d/a$ (see Fig.~\ref{fig:fig1}a) as a function of the junction angle $\alpha$. 
($\bullet$) : experimental measurements; ($+$) : results of the numerical simulations.}
\end{figure}
More quantitatively,  the distribution of each fluid between
the two outlets will be characterized by the volume fraction $q$ occupied
 by the fluid of interest: practically $q$ is taken to be equal to $a_d/a$ in which 
 $a$ is the total width of the section and $a_d$ the width of the flow tube corresponding
 to this fluid in the parallel flow region. Here, $a_d$ is computed for the dyed solution  in  outlet $2$
where its   volume fraction is lower (Figs.~\ref{fig:fig0} and \ref{fig:fig1}a).
 Several measurements of $a_d$   in the parallel flow region are averaged in order to compute $q$. 
Figure~\ref{fig:fig2} displays the experimental variation ($\bullet$ symbols) of the fraction $q$ with  $\alpha$:
$q$ is exactly zero below a threshold angle $\alpha_c$ of experimental value in the range
$30^\circ  \le \alpha_c \le 35^\circ$. Above $\alpha_c$, $q$ increases linearly with $\alpha$ 
up to $q = 0.5$ for $\alpha = 90^\circ$.

The zero value of  $q$ below  the critical  angle  $\alpha_c$  may seem at first counter-intuitive: 
it implies indeed that the fluid particles 
follow preferentially,  in the region of 
the junction, the streamlines with the highest curvature.  
Actually, this reflects  a key characteristic of all  viscous Stokes flows, namely that 
they must achieve a minimum viscous energy dissipation.
Let us assume that,  in contrast to the experiment of Fig.~\ref{fig:fig1}c,  the dyed fluid flows into outlet $1$ 
and the transparent one into $2$ at low values of $\alpha$. This requires a counterflow of the 
two fluids along segment $CD$ which is  much longer than $AB$; moreover,
this small  length of $AB$  increases the transverse velocity gradient. 
The viscous energy dissipation, proportional to the  
integral of the square of this gradient, would then be too high. In the experimental flow of  
 Fig.~\ref{fig:fig1}c,  instead, the distance  $AB$ along which the two flows  coexist is smaller
 than the  transverse distance $CD$: together with  the appearance of  recirculation cells with very low velocities (and  gradients), this leads, indeed, to a lower dissipation than in the first case.
The result would  be completely different for the inertial potential flow of a perfect fluid  in the same geometry.  Let us consider, for instance, the comparable classical problem of a $2D$ jet of perfect fluid impinging obliquely at an angle $\alpha$ on a plane wall~\cite{textbook}: then, as $\alpha$ decreases from $90^\circ$ to zero, 
the fraction (equivalent to $q$) of the fluid following the path with the lowest curvature increases from 
$0.5$ to $1$ (instead of decreasing to $0$ like in Fig.~\ref{fig:fig2}).
The structure of the flow in the center part of the junction is obviously a key element for predicting the exchanges
between the two flowing fluids.
Additional information is  obtained by saturating  the intersection with the rhodamin-water
solution instead of pure water before establishing a stationary flow in the facing injection channels.
No difference is observed for $\alpha > \alpha_c$ but, for $\alpha < \alpha_c$, the fluorescence
light intensity is higher in the center of the junction than when it was initially saturated by 
transparent fluid. This suggests that some dye is trapped in the center of the junction and
that the latter corresponds to a dead or recirculation zone. 

\begin{figure}[htbp]
\includegraphics[width=9.5cm]{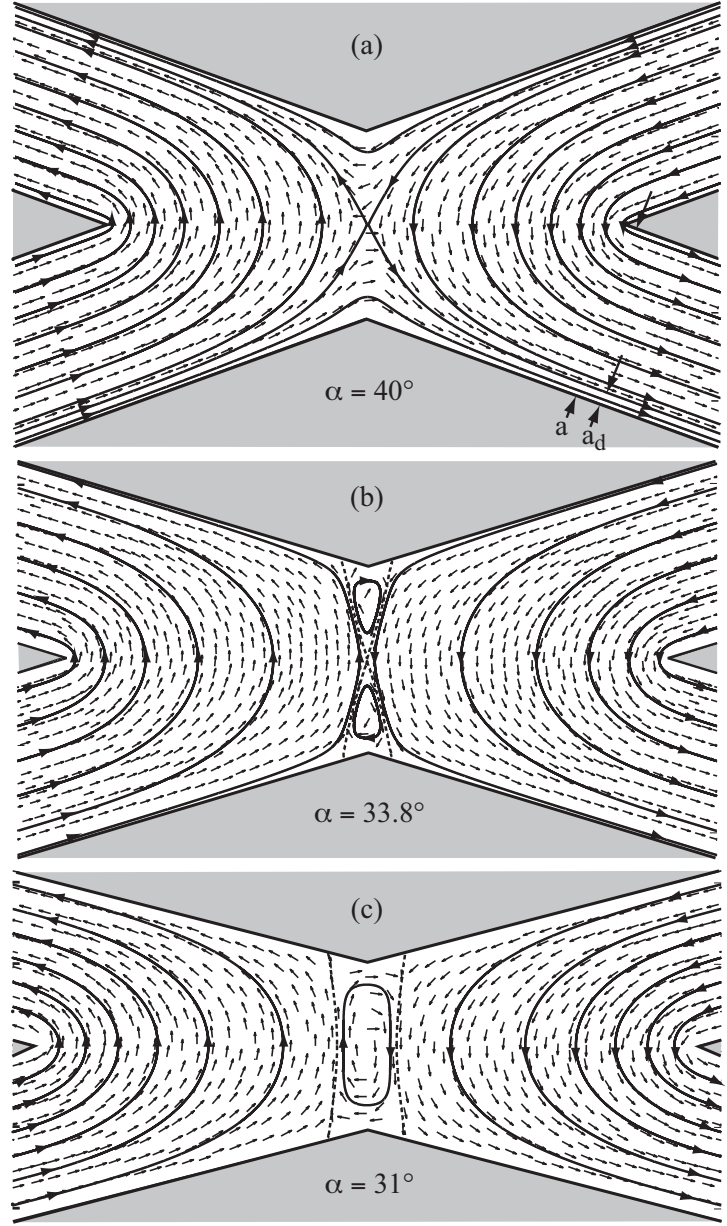}
\caption{\label{fig:fig3}Velocity field (vectors) obtained numerically for : (a) $\alpha = 40^\circ$; (b)  $\alpha = \alpha_c = 33.8^{\circ}$ (c) $\alpha = 31^\circ$. The length of all vectors are equal for a better visibility so that only the orientation of the velocity is represented. Continuous lines represent selected streamlines of the flow; dashed lines mark the boundaries of the flow recirculation regions.}
\end{figure}
In order to verify this assumption, the $2D$ Stokes equation  has been  solved numerically in the $(x,y)$ plane by means of the FreeFem finite elements package~\cite{Freefem2010} : the lack of inertial effects and the $2D$ distribution of the two fluids suggest indeed that this equation may predict adequately the distribution of the two fluids. 
\cor{The velocity vector field and the sample streamlines obtained in this way  for 
three values of the angle $\alpha$ are displayed in Fig.~\ref{fig:fig3}.} 
For $\alpha = 31^\circ$ (Fig.~\ref{fig:fig3}c), a recirculation cell is observed in the center of the junction as suggested by the experiments. It separates completely the flows on the two sides of the junction. The boundary of this zone inside which the flow lines are not connected to the outside is marked by  dashed lines. The modulus of the velocity (not represented on the figure) is about $100$ times lower inside the  cell than outside.
The detailed structure of the recirculation flow depends on the angle $\alpha$ and will be discussed below.
For $\alpha = 40^\circ$ (Fig.~\ref{fig:fig3}a), there is only one stagnation point and it is located at the center of symmetry of the junction: streamlines originating at this point separate those extending across the junction from those which ``bounce back'' on it.  The distance of these particular streamlines to the nearest wall represents therefore the distance $a_d$ defined above. This provides a numerical prediction of the ratio $q = a_d/a$: its variation as a function of the angle $\alpha$ is displayed 
by $(+)$ symbols in Fig.~\ref{fig:fig2}. The numerical values of $q$ are first very close to the experimental ones:  this supports ({\it a posteriori})  the $2D$ approximation. The  variation of $q$ with $\alpha$ is observed to be very  linear except close to the critical angle $\alpha_c$: this confirms the trend of the experimental data.
\cor{These  simulations  provide the precise value $\alpha_c = 33.8 \pm 0.2^\circ$. For $\alpha = \alpha_c$ (Fig.~\ref{fig:fig3}b), one observes two recirculation cells  on the lateral edges extending up to the central stagnation point. For this and lower angles, there is no flow path connecting the left and right sides of the junction any longer.}

\begin{figure}
\includegraphics[width=\W]{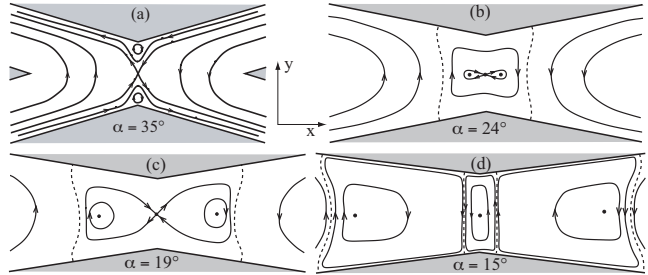}
\caption{\label{fig:fig4}\cor{Typical streamlines obtained numerically for different angles (a): $\alpha=35^\circ$), (b): $\alpha=24^\circ$, (c): $\alpha= 19^\circ$, (d): $\alpha= 15^\circ$. Continuous and dashed lines have the same meaning as in Fig.~\ref{fig:fig3}}}
\end{figure}
The recirculation zones observed at low angles $\alpha$ clearly play a major part in the exchange of passive scalars (solutes, heat $\ldots$) between the injected flows. Their structure depends strongly on $\alpha$ as shown in 
Figure~\ref{fig:fig4} displaying typical examples obtained  numerically. 
\cor{For $\alpha$ slightly above $\alpha_c$  (Fig.~\ref{fig:fig4}a), side recirculation cells are present, like for $\alpha = \alpha_c$: they reduce   the section available in the plane of symmetry  (vertical on the figure) for streamlines crossing from one side  to the other but do not block it.  This section becomes zero for $\alpha = \alpha_c$ as the two cells touch each other  (Fig.~\ref{fig:fig3}b) and then, below $\alpha_c$, a single cell extending all across the junction appears  (Fig.~\ref{fig:fig3}c). The faster  variation of $q$ near  $\alpha_c$ in Fig.~\ref{fig:fig2} may reflect the scarcity of the  flow paths \cor{connecting the two sides of the junction}  and the fact that they  cross low velocity regions.
Decreasing further $\alpha$ (Fig.~\ref{fig:fig4}b)} results in an elongation of the cell in the $x$ direction with, inside it, two smaller cells joined at a stagnation point (Fig.~\ref{fig:fig4}b). At a still lower angle, the latter occupy nearly the full width of the channel (Fig.~\ref{fig:fig4}c) until an additional  cell, elongated in the $y$ direction, appears in the center (Fig.~\ref{fig:fig4}d). The typical velocity decreases by a large factor ($\simeq 100$) between the outer flow and the first generation of cells and between the latter and the next generation. New cells appear at still lower angles (not shown on the figure) and correspond to vanishingly small velocities.

Vortices have already been observed in Stokes flow by many other authors. A classical example is Moffatt vortices induced  by an outside flow  in a corner of angle $\beta < 146^\circ$ between plane walls: an infinite sequence of eddies appears with velocities decreasing strongly, like here, from one generation to the next~\cite{Moffatt1964,Jeffrey1980}. 
Here, a similar  sequence appears at low angles $\alpha$ but is interrupted at the middle of the junction with a symmetrical sequence starting on the other side.
These Stokes flows in corners do not display however like in  Figs.~\ref{fig:fig4}a-c pairs of corotating vortices separated by a stagnation point unless a planar Couette flow is superimposed~\cite{Wilson2005}. These latter structures are also observed  in  elongated rectangular cavities when the two smaller parallel walls are sheared in opposite directions~\cite{Gurkan2003}. Like at low $\alpha$ values in the present system, as the elongation of the cavity increases, new recirculation cells appear : one observes first a pair of vortices and they  coalesce into a single cell when the elongation is increased further.

The experiments and simulations  presented here demonstrate that, at all  $\alpha$  values (except 
$90^\circ$), the largest part of (if not all) each injected flow bounces back  into the outlet at the  angle 
$\alpha$ to the injection.  The actual fraction depends strongly on $\alpha$, as shown above, but will  also vary 
with the aspect ratio $h/a$ of the section: for instance, if $h/a \ll 1$ (instead of $h/a = 1$), one obtains a 
 Hele Shaw cell geometry. Then, the average of the velocity field  over the depth $h$  would be  similar to that  
 for a perfect fluid and the variation of $q$ with $\alpha$ will be completely different.  
 It will be important to study the transition between these two 
opposite behaviors by varying continuously $h/a$. Since the flow field is controlled by the viscous dissipation, it 
is likely to be strongly influenced by the rheological properties of the fluid. Moreover,  anomalies are known to occur for fluids of high extensional viscosity in the vicinity of the stagnation points in different flow geometries~\cite{Perera82,Chow88,Harlen90,Harris93}. For viscoelastic fluids flowing in orthogonal channels, elastic instabilities have indeed been observed experimentally~\cite{Arratia2006} and flow asymmetries have been predicted numerically~\cite{Poole2007}.

The recirculation cells appearing at low $\alpha$ values will  control the exchange of solute (or other passive or reactive species)  between the two injected fluids: this exchange involves a combination of transverse molecular diffusion  between the cells and the two flows (and from one cell to another) and  convective transport by the circulation within the cells. 
For reactive fluids, the recirculation cells may act as microreactors into which species contained  in the injected fluids are transferred through transverse molecular diffusion.  It may be possible to induce a more efficient chaotic mixing in the  recirculation cell by varying  the geometry of the intersection  and/or by using suitable time variations of the relative values of the two injection flow rates.
\begin{acknowledgments}
We thank R. Pidoux, L. Auffay and A. Aubertin for realizing and developing the experimental set up, \cor{D. Etien for his contribution to the experiments} and D. Salin, E.J. Hinch and H.A. Stone for illuminating suggestions.
We acknowledge the support of the RTRA ``Triangle de la Physique'' and of the LIA PMF-FMF (Franco-Argentinian International Associated Laboratory in the Physics and Mechanics of Fluids).
\end{acknowledgments}
\bibliography{apssamp}

\end{document}